\documentclass[aps,prb,twocolumn,10pt,amsmath,amssymb,showpacs]{revtex4-1}
\usepackage{graphicx}
\usepackage{dcolumn}
\usepackage{bm}
\usepackage{longtable}

\begin{document}
\title{Capillary condensation in a square geometry with surface fields}
\author{M.~Zubaszewska,$^{1}$ A.~Gendiar,$^{2}$ and A.~Drzewi\'{n}ski$^{1}$}
\affiliation{$^{1}$Institute of Physics, University of Zielona G\'ora, ul. Prof. Z. Szafrana 4a,
65-516 Zielona G\' ora, Poland}
\affiliation{$^{2}$Institute of Physics, Slovak Academy of Sciences, SK-845 11 Bratislava, Slovakia}

\date{\today}

\begin{abstract}
We study the influence of wetting on the capillary condensation for a simple fluid 
in a square geometry with surface fields, where the reference system is an infinitely long slit.
The Corner Transfer Matrix Renormalization Group method has been extended to study
a two-dimensional Ising model confined in an $L \times L$ geometry with equal surface fields.
Our results have confirmed that in both geometries the coexistence line shift is governed by
the same scaling powers, but their prefactors are different.
\end{abstract}


\maketitle

Porous materials are solids consisting of interconnected network of pores. In recent years, microporous and
mesoporous materials have been a focus of nanoscience and nanotechnology as their properties differ
significantly from the same bulk materials \cite{Israelachvili,Sattler}. Both the pore size and its shape,
as well as the chemical nature of its surface, i.e., whether hydrophilic or hydrophobic, determine
the properties of porous materials \cite{Gelb}.
In a mesopore, the cumulative effect of the walls becomes important. After the formation of adsorbate layers with
2-3 molecular thickness on walls, further adsorption induces attractions between adsorbate molecules, leading to
a sudden condensation of liquid-like adsorbate molecules inside the pores. This effect is analogous to
the capillary condensation phenomena \cite{Evans}.

The Ising square of a finite size $L$ with the field $h_1$ acting on all four surfaces (boundaries) can be used as
an idealized representation of a simple fluid in a pore or between finely divided colloidal particles. 
The choice of the model entails the assumption that the intermolecular forces are short-ranged in character,
there are no dispersion-like forces. In fluid experiments the equivalent
of the bulk magnetic field $H$ describes the deviation of the critical chemical potential $H \sim \mu - \mu_0$
that is determined by the density of the fluid in the reservoir. The phenomenon equivalent to the capillary condensation
can be studied in magnetic systems \cite{Binder}. Generally, in the bulk, the phase coexistence occurs for
temperatures $T<T_c$ and for vanishing bulk magnetic field $H$. In a slit $L \times  \infty$ with identical surface
fields at the boundaries, the combined effect of surface fields and confinement shifts the phase coexistence
to a non-zero value of the bulk magnetic field $ H=H_{coe}(L) $, which for large $L$ scales as

\begin{equation} \label{e1}
H_{coe} (L) = \frac{\sigma_0 \cos\theta}{m_b} \frac{1}{L},
\end{equation}
where $\sigma_0$, $m_b$, and $\theta$ are the surface tension of the free up-spin/down-spin interface, bulk
spontaneous magnetization, and contact angle given by Young's equation, respectively \cite{Fisher}. In our
case the liquid phase is represented by up-spins, gas phase by down-spins, whereas area without spins is
identified with the wall. The above equation is known in the literature as the Kelvin equation \cite{Kelvin}.

\vspace{5mm}
\begin{figure}[tbh]
\centering
\includegraphics[scale=0.32]{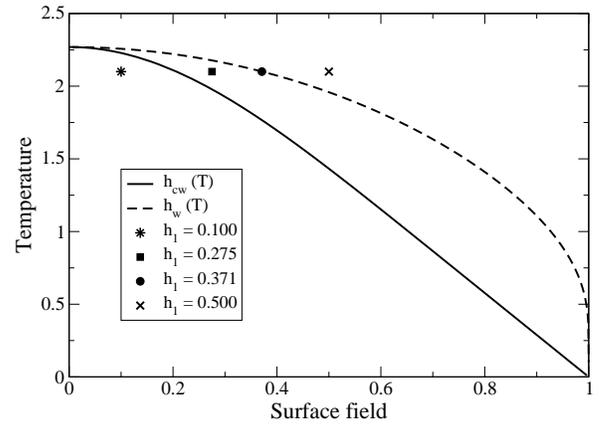}
\caption{The critical wetting lines for semi-infinite systems at $T=2.1$: the corner wetting ($h_{cw}(T)$)
on the two planar surfaces forming a straight angle and wetting ($h_{w}(T)$) on the planar surface.
With respect to the planar wetting $h_1=0.1$ and $0.275$ values correspond to the dry regime, whereas
the $h_1=0.371$ and $0.5$ value to the wet one.}
\label{fig:01}
\end{figure}

Wetting occurs in systems close to the line of phase coexistence when one phase
may adsorb preferentially at a solid substrate. Typically, we model the substrate using a planar surface
where the critical line $h_{w}(T)$, presented in Fig.~(\ref{fig:01}), is known exactly \cite{Abraham}

\begin{multline}
\label{e21}
\exp(2 J/k_BT) \left( \cosh(2 J/k_BT) - \cosh(2 h_{w} J/k_BT) \right)\\
= \sinh(2 J/k_BT).
\end{multline}

For this semi-infinite system if we approach the coexistence line from the gas side along a given isotherm
($T_c > T > T_w$) the amount of liquid adsorbed on the surface $l$ diverges $l \sim H^{-\beta_s^{co}}$.
Moreover, the phase transition, called the complete wetting, is characterized by the presence of the singular part
of the excess surface free-energy $f_{sing} =  H^{2-\alpha_s^{co}}$.
For the two-dimensional Ising model, the values of the critical exponents are $2-\alpha_s^{co} = 2/3$ and
$\beta_s^{co}= 1/3$.

Consequently, when we consider the (pseudo)two-dimensional Ising system in a slit geometry
and the interaction between the liquid and walls is strong (above the $h_{w}(T)$ line), the Kelvin equation
fails and some corrections are necessary. The reason is that, although the macroscopically thick layer forms only
for a semi-infinite system, a noticeable liquid layer intervenes between a gas and the wall for a finite-size system
as well.
Therefore, Foster pointed out \cite{Foster} that if adsorbed layers were formed prior to condensation, the slit
width $L$ in Eq.~(\ref{e1}) should be corrected by the layer thickness $l$. Derjaguin showed \cite{Derjaguin} that
if solid-fluid forces decayed exponentially or had a finite range, the pore width $L$ could be replaced by $L - 2l$.

Next, Evans et al. \cite{Evans} showed that the effects of wetting layers were of quantitative rather
than of qualitative importance for capillary condensation. Albano et al. \cite{Albano} and Parry and
Evans \cite{Parry} analyzed the next-order correction term to the Kelvin equation for the semi-infinite system. Both
studies, using scaling and thermodynamics arguments, concluded that for temperatures below
the wetting temperature $T_w$ (the dry regime) the leading correction to scaling term was of type $L^{-2}$.
Above the wetting (the wet regime) the correction is expected to be non-analytic due to a singularity of
the surface free energy. For the two-dimensional Ising model \cite{Albano,Parry} the predicted correction term is proportional to $L^{-5/3}$.

In the subsequent numerical investigation, the density-matrix renormalization techniques were
employed \cite{Carlon,Szota}. For a large range of surface fields and temperature higher-order
corrections were not compatible with $L^{-5/3}$, but they were of type $L^{-4/3}$. It has been shown that this
apparent disagreement was due to the fact that even for the large sizes considered ($L\sim150$) the wetting layer
has a limited thickness, so that the singular part of the surface free energy that determines the correction-to-scaling
behavior is dominated by the contacts with the walls.

In real systems, where properties of both pure fluids and fluid mixtures confined to nanoporous and microporous
materials are under consideration \cite{Brov}, the square-like geometry is more common than the slit one. In 
this case, the geometry of the system significantly affects the course of wetting phenomena because close to
a corner the impact of the individual walls is strong, which should lead to more intensive formation of the wetting
layer. Therefore, the corner wetting transition should also be taken into account \cite{Parry2,Milchev}. As two
sides of the square form the straight angle, the corresponding ($L \to \infty$) corner wetting line $h_{cw}(T)$,
presented in Fig.~(\ref{fig:01}), is known exactly \cite{Wood,Maciolek}

\begin{multline}
\label{e22}
\cosh(2 J/k_BT) - \exp(-2 J/k_BT) \sinh^2(2 J/k_BT)\\
=\cosh(2 J h_{cw}/k_BT).
\end{multline}

In order to model the influence of wetting phenomena on the capillary condensation in a square geometry we consider
a square Ising ferromagnet subject to identical boundary fields with the following Hamiltonian:
\begin{equation} \label{e23}
{\cal H} = -J \left( \sum_{ijk\ell} S_{i,j} S_{k,\ell} - h_1 \sum\limits_{{\rm surface}\atop{\rm spins}} S_{i,j} -
H \sum\limits_{{\rm all}\atop{\rm spins}}  S_{i,j} \right),
\end{equation}
with $J>0$ and $S_{i,j} = \pm 1$. The first sum is taken over the nearest neighbours, while the second sum is
performed on spins
at the surface only. The surface field $h_1$ corresponding to direct short-range interactions between the walls
and spins is related to the preferential adsorption on the surface for one of the two phases. The uniform bulk
magnetic field $H$ acts over all spins.

\vspace{5mm}
\begin{figure}[tbh]
\centering
\includegraphics[scale=0.28]{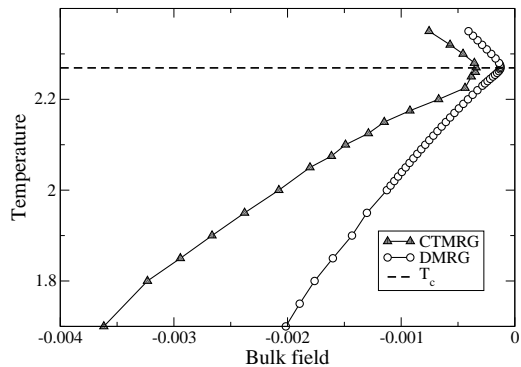}
\caption{Pseudo-coexistence lines for the square (triangles) and slit (circles) systems for $L=500$ and $h_1=0.8$.
$L$ is measured in units of lattice constant, whereas $h_1$ is in $J$ units.}
\label{fig:02}
\end{figure}

The origin of applied numerical method, called the Corner Transfer Matrix Renormalization Group (CTMRG), came from
Baxter \cite{Baxter}. Next Nishino and Okunishi \cite{Okunishi} combined his corner transfer matrix method with
the ideas from the density matrix renormalization group method (DMRG) approach. The last technique was developed
by White \cite{White,Uli} for the study of ground state properties of quantum spin chains and next extended by
Nishino to two-dimensional classical systems in a slit geometry \cite{Nishino}.

The general idea is to find a representation of the configurational space in a restricted space which is much
smaller than the original one $m \ll 2^{L^2}$. This truncation is done through the construction of a reduced density
matrix whose eigenstates provide the optimal basis set $m$. Of course, the larger $m$, the better accuracy, so in
the present case we keep this parameter up to $m = 400$. 

Although in the original CTMRG algorithm the full transfer matrix is never constructed, we have modified it
to determine the two eigenvectors related to the largest eigenvalues. Because each of these vectors dominates
on the opposite side of the coexistence line, using both vectors for the construction of the density matrix
guarantees that the Hamiltonian is properly projected on the subspace of most probable states. Our results have not
shown any ambiguities of the calculated free energy and by increasing the number of states kept $m$, our results
(the free energy) converged. To have a point of reference, we compared the results for the square geometry with
the results for the slit geometry, where the DMRG technique was applied \cite{Carlon,Szota}.

Due to the finiteness of $L$ and to the nonvanishing surface field $h_1$ the (pseudo)coexistence lines
(for both geometries) are shifted with
respect to the bulk coexistence line ($H=0$). As one can see in Fig.~(\ref{fig:02}) this effect is much
stronger for the square-like system. This is easy to understand, if one remembers that the influence
of walls on the individual spins is here reinforced with respect to the slit geometry.

\begin{figure}[tbh]
\centering
\includegraphics[width=0.48\textwidth,clip]{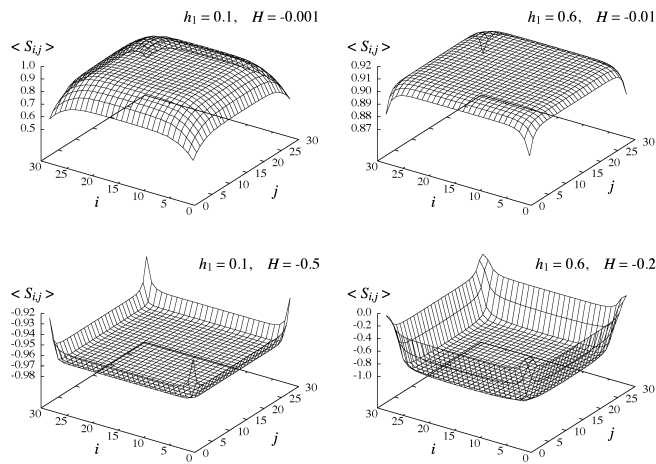}
\caption{Two-dimensional magnetization profiles $\langle s_{i,j}\rangle$ on the square geometry
$31\times31$ for the two values of the surface field $h_1=0.1$ (left graphs) and $h_1=0.6$ (right graphs)
calculated at fixed temperature $T=2.0$. The profiles change their shapes and spin polarizations
while varying the bulk field value $H$ below and above the wetting temperature
$T_w(h_1=0.1)=2.257$ and $T_w(h_1=0.6)$=1.814, respectively.}
\label{fig:02a}
\end{figure}

A schematic draw of the fluid shape adopted in a square geometry below ($h_1=0.1$) and above ($h_1=0.6$)
wetting is demonstrated in Fig.~(\ref{fig:02a}). Both upper magnetization profiles correspond to the phase
when a liquid fills the square-like pore. For a weak field the magnetization is lower nearby walls, whereas
for a strong field the magnetization is essentially uniformly high throughout the square. The bottom
magnetizations illustrate the situation on the other side of the coexistence line, when the gas phase occupies
the middle of the square. While the weak-field magnetization profile only slightly increases at the square edges, 
the strong-field magnetization profile increases considerably, which corresponds to creation of the liquid layer.
Note that the surface impact is always enhanced in the square corners, although limited to a relatively small area.

\vspace{5mm}

The real phase transition (the complete wetting) occurs only in a semi-infinite system. Since we deal with
a finite-size system, the thickness of the wetting layer is limited, and a sharp liquid-gas interface is not
observed. Furthermore, because for a weak surface field (below the wetting line) there remains a thin
liquid layer between a gas and the wall, the gas-wall interface formally never occurs. For the same reason,
below the wetting line the isolated droplets of liquid cannot be observed on the wall, so the contact angle cannot
be drawn.

As a useful tool for the analysis of the higher terms of the Kelvin equation, we introduce the logarithmic
derivative which acts as an effective dominant exponent
\begin{equation} \label{e3}
\alpha (L) = -\frac{\ln[H_{coe}(L+\Delta L)]-\ln[H_{coe}(L)]}{\ln (L+\Delta L) -\ln (L)}.
\end{equation}
When the following expansion of the Kelvin equation is assumed

\begin{equation} \label{e4}
H_{coe} = \frac{A}{L^{\alpha}} + \frac{B}{L^{\gamma}},
\end{equation}
combining both formulae, we obtain the first-order expansion for the effective exponent

\begin{equation} \label{e33}
\alpha (L) = \alpha + (\gamma - \alpha) \frac{B}{A} \frac{1}{L^{\gamma - \alpha}}.
\end{equation}

The Kelvin equation is expected to be valid to the first order for all $T<T_c$, but since only a limited size $L$
is available for the numerical computation, it is preferable to consider only temperatures not too close to
the bulk critical temperature, where the scaling of the capillary critical point $H_{crit}(L) \sim L^{15/8}$ is
present \cite{Fisher}. Therefore, our results refer to temperature $T = 2.1$, but we have checked that they do not qualitatively change for other temperatures.

\vspace{5mm}
\begin{figure}[tbh]
\centering
\includegraphics[scale=0.3,clip]{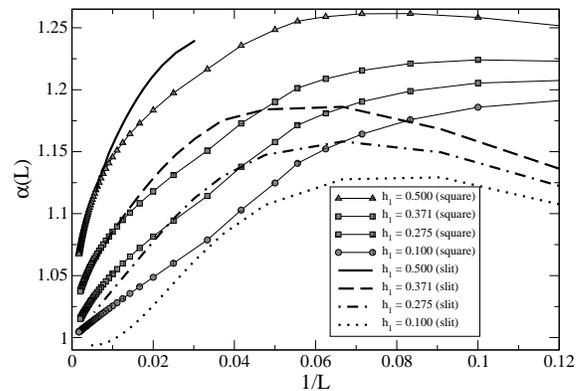}
\caption{Plots of the local exponent for both geometries at $T=2.1$. Lines correspond to the square geometry,
whereas the symbols are related to the slit one.}
\label{fig:03}
\end{figure}

Figure~(\ref{fig:03}) shows that for a wide range of the surface fields, when the system grows,
the value $\alpha (L)$ goes to $\alpha =1$. It confirms that to the accuracy of the first-order expansion,
the shift of the phase coexistence and the system size are reciprocal in both geometries.

For the wet phase ($\theta =0$) in the slit geometry, the exact formula for the coefficient is
known (see Eq.(\ref{e1})) giving $A(T=2.1)=0.335$ which very well agrees with our estimated value $0.345$. Generally,
our numerical fittings for curves $H_{coe} (L)$ according to Eq.(\ref{e4}) show that the ratio of the main
prefactors for the slit and square geometries is $A_{slit} / A_{square} \sim 1/2$ for each value of the surface
field. For example, below the wetting $A_{slit}(h_1 = 0.1)=0.121$ and $A_{square}(h_1 = 0.1)=0.242$, whereas above
the wetting $A_{slit}(h_1 = 0.5)=0.345$ and $A_{square}(h_1 = 0.5)=0.669$.  This fact can be explained in the following
way: when the size of the square increases, the impact of the corners remains more or less the same, but the influence
of the sides increases. Thus, when $L$ goes to infinity, the system begins to resemble a set of two slits perpendicular
to each other.

\vspace{5mm}
\begin{figure}[tbh]
\centering
\includegraphics[scale=0.3,clip]{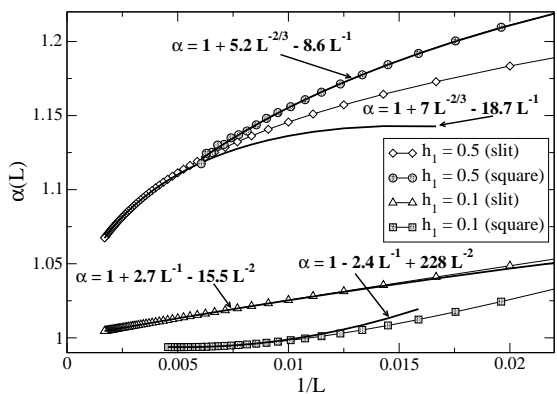}
\caption{The effective exponents $\alpha (L)$ and their fitting curves denoted by the thick lines for two sample
values of the surface field: $h_1 = 0.1$ for the dry system and $h_1 = 0.5$ for the wet
system (see Fig.~(\ref{fig:01})).} 
\label{fig:04}
\end{figure}

As one can see in Figs.~\ref{fig:03} and \ref{fig:04}, the shape of the curves indicates that we are dealing with
various expressions of a higher order. First of all, the $B$ coefficient that is always positive in the slit
geometry, is negative in the dry regime for the square geometry. This is manifested by the $\alpha (L)$ function
minima for small $1/L$, where the leading correction $B/L^{\gamma}$ dominates. Obviously, the next correction has
to be positive. Moreover, in the wet regime, where the $B$ coefficients are positive, as one can see
in Fig.~(\ref{fig:03}), the curves corresponding to the same value of the surface field start to overlap when $L$
becomes enough large.

In the square geometry, as far as the leading correction terms are concerned, the precise analysis of our numerical
results shows, in both the dry and wet regimes, they are the same as in the slit geometry (see Fig.~\ref{fig:04}).

It is worth adding that the value of the surface field $h_1 = 0.1$ ($h_1 = 0.5$) was chosen to be
in the dry (wet) regime with respect to both wetting curves (see Fig.~\ref{fig:01}). Although the presented
curves correspond to infinite systems, we are dealing with systems large enough to guarantee that both points
are localised on the appropriate side of the wetting curves.

To sum up, the Corner Transfer Matrix Renormalization Group method has been extended to study the equilibrium
statistical mechanics of simple fluids confined in the square geometry providing some input for
understanding experiments in porous materials. As the method is not perturbative, it can be applied to arbitrary
values of the model parameters and yields accurate results, provided that a convergence of the free energy is
reached. Accuracy of the CTMRG method is completely controlled by varying the number of the states kept.
Moreover, the method does not suffer from getting stuck in a local minimum of the free energy instead of the true
global free energy minimum what is a typical problem in Monte Carlo simulations. There are neither metastabilities
nor hysteresis effects provided that $m$ is set to be sufficiently large which is our case concluded by the converged
free energy.

Our results confirm that in a more realistic (square) geometry the coexistence line shift is inversely
proportional to the system size $L$ and the main pre-factor is two times larger than for the case of
an infinitely long slit of width $L$. Moreover, for the square geometry the leading correction term to
the Kelvin equation becomes negative in the dry regime.

We have also found that similarly to the slit geometry the leading corrections to scaling in the square geometry
are of type $1 / L^2$ in the dry regime and of type $1/ L^{5/3}$ in the wet regime. The last correction is
again non-analytic due to a singularity of the surface free energy.

Curve fitting was carried out for the values of the surface fields that correspond to the dry/wet regime
with respect to both wetting lines defined by Eqs.~\ref{e21} and \ref{e22}. The area between the wetting
lines was not analysed in detail, although the strongest competition can be expected between the two
types of wetting. This requires an examination of larger systems, which
implies the need for a more precise calculation. In our next studies, we aim to increase the precision of
the numerical results by additional improvement of the CTMRG algorithm.

\section{Acknowledgements}
This work was done under projects VEGA-2/0074/12, APVV-0646010 (COQI) and POKL.04.01.01-00-041/09-00.
Numerical calculations were performed in WCSS Wroc\l aw (Poland, grant 82).

\end{document}